\documentclass[12pt]{article}
\usepackage{amsmath}

\usepackage{amsthm}
\usepackage{amssymb}
\usepackage{stmaryrd}
\usepackage{subfigure}
\usepackage{empheq}
\usepackage{xcolor}
\usepackage{float}
\usepackage[demo]{graphicx}
\renewcommand{\theequation}{\arabic{equation}}
\newcommand{\EQ}{\begin{equation}}
\newcommand{\EN}{\end{equation}}

\newcommand{\ket}[1]{\left|#1\right\rangle}      % Ket-Zustand
\newcommand{\bear}{\begin{eqnarray}}
\newcommand{\ear}{\end{eqnarray}}
\newcommand{\bt} { \begin{tabular} }
\newcommand{\et}{ \end{tabular} }
\newcommand{\bc} { \begin{center} }
\newcommand{\ec}{ \end{center} }

\newcommand{\btb} { \begin{table} }
\newcommand{\etb}{ \end{table} }
\begin{document}

\topmargin 0pt
\oddsidemargin 5mm
\newcommand{\NP}[1]{Nucl.\ Phys.\ {\bf #1}}
\newcommand{\PL}[1]{Phys.\ Lett.\ {\bf #1}}
\newcommand{\NC}[1]{Nuovo Cimento {\bf #1}}
\newcommand{\CMP}[1]{Comm.\ Math.\ Phys.\ {\bf #1}}
\newcommand{\PR}[1]{Phys.\ Rev.\ {\bf #1}}
\newcommand{\PRL}[1]{Phys.\ Rev.\ Lett.\ {\bf #1}}
\newcommand{\MPL}[1]{Mod.\ Phys.\ Lett.\ {\bf #1}}
\newcommand{\JETP}[1]{Sov.\ Phys.\ JETP {\bf #1}}
\newcommand{\TMP}[1]{Teor.\ Mat.\ Fiz.\ {\bf #1}}

\renewcommand{\thefootnote}{\fnsymbol{footnote}}

\newpage
\setcounter{page}{0}
\begin{titlepage}
\begin{flushright}
\end{flushright}
%\vspace{0.2cm}
\begin{center}
{\large  Bethe equations for the critical three-state Potts spin chain with toroidal boundary conditions} \\
\vspace{0.5cm}
{\large M.J. Martins } \\
\vspace{0.15cm}
{\em Universidade Federal de S\~ao Carlos\\
Departamento de F\'{\i}sica \\
C.P. 676, 13565-905, S\~ao Carlos (SP), Brazil\\}
%E-mail Address: {\tt martins@df.ufscar.br}}\\
\vspace{0.35cm}
\end{center}
%\vspace{0.5cm}
%\centerline {In honor to Prof.R.J. Baxter on the occasion of his 75th birthday}
%\vspace{0.1cm}
\begin{abstract}
In this paper, we consider the parameterization of the spectra of the three-state 
critical Potts quantum chain with integrable twisted boundary conditions in terms 
of Bethe ansatz type equations. The Bethe equations are found by investigating 
the structure
of the eigenvalues of the respective twisted transfer matrices, and with the help
of certain identities satisfied by the product of transfer matrices 
operators. We have studied the  
completeness of the spectrum in terms of the Bethe roots for small lattice sizes
and have computed the eigenstate momenta. We found that the spins
of the low-lying excitations
can have fractional values in accordance with 
predictions of the 
underlying conformal 
field theory. We argue that our framework can be used to build 
integrable Hamiltonians whose spectra are determined by mixing
different toroidal boundary conditions.

\end{abstract}
%\vspace{.15cm} 
%\centerline{}
%\vspace{.1cm} 
\centerline{Keywords: three-state Potts quantum chain, twisted boundaries, Bethe equations. }
%\vspace{.15cm} 
\centerline{January~~2026}
\end{titlepage}

%\tableofcontents

\pagestyle{empty}

\newpage

\pagestyle{plain}
\pagenumbering{arabic}

\renewcommand{\thefootnote}{\arabic{footnote}}
\newtheorem{proposition}{Proposition}
\newtheorem{pr}{Proposition}
\newtheorem{remark}{Remark}
\newtheorem{re}{Remark}
\newtheorem{theorem}{Theorem}
\newtheorem{theo}{Theorem}

\def\ll{\left\lgroup}
\def\rr{\right\rgroup}

\newtheorem{Theorem}{Theorem}[section]
\newtheorem{Corollary}[Theorem]{Corollary}
\newtheorem{Proposition}[Theorem]{Proposition}
\newtheorem{Conjecture}[Theorem]{Conjecture}
\newtheorem{Lemma}[Theorem]{Lemma}
\newtheorem{Example}[Theorem]{Example}
\newtheorem{Note}[Theorem]{Note}
\newtheorem{Definition}[Theorem]{Definition}

\section{Introduction}

In general, integrable classical lattice spin models of 
statistical mechanics
are usually considered by imposing periodic boundary conditions on the 
two-dimensional lattice \cite{BAX}. However, it is expected that 
integrability will remain intact for the other boundary
conditions compatible with the torus and the system should be
solvable under more generic twisted boundary conditions.
The typical example is the $Z(2)$ invariant Ising model
which is considered to be exactly solvable for both periodic 
and antiperiodic 
boundary conditions \cite{MCC}. In this case,  the antiperiodic boundary 
is equivalent
of changing the sign of the exchange interaction at one end of the
toroidal lattice.
The purpose of this paper is to consider this problem for one of the
simplest generalization of the Ising model, where the spin variables
can assume three possible values. This is the so-called
three-state scalar Potts model in which 
the interactions among adjacent 
spins are assumed to have the same thermal energy \cite{POT}. 
On the square lattice, this system has only
two horizontal and vertical edge weights, and it becomes solvable 
when such weights
sit on a curve in which the model is self-dual \cite{BAX1}.  We recall that such a
manifold can be regarded as a special case of the $n$-state integrable
$Z(n)$ invariant spin model discovered by Fateev and Zamolodchikov \cite{FAZA}.  
It turns out that the   
horizontal $W_h(x)$ and vertical $W_v(x)$  
edge weights of the integrable three-state scalar Potts model have 
the following structure \cite{FAZA},
\begin{equation}
W_h(x)=\left(
\begin{array}{ccc}
1 & a(x) & a(x)  \\
a(x) & 1 & a(x) \\ 
a(x) & a(x) & 1 \\ 
\end{array}
\right),~~~ 
W_v(x)=\left(
\begin{array}{ccc}
1 & b(x) & b(x)  \\
b(x) & 1 & b(x) \\ 
b(x) & b(x) & 1 \\ 
\end{array}
\right) , 
\label{weigh}
\end{equation}
where the variable $x$ parameterizes the self-dual curve. 
The non-trivial matrix elements
$a(x)$ and $b(x)$ are determined by,
\begin{equation}
a(x)=\frac{\sin(\frac{\pi}{6}-x)}{\sin(\frac{\pi}{6}+x)},~~~
b(x)=\frac{\sin(x)}{\sin(\frac{\pi}{3}-x)} .
\label{pesos}
\end{equation}

One consequence of integrability is that we can 
construct a family of commuting transfer matrices, and from that we derive 
the underlying 
quantum spin chain \cite{BAX}. In the case of classical 
spin models, the suitable 
transfer matrix is built
by combining the horizontal and vertical edge weights 
along the diagonals of a square $L \times L$ lattice following
the original construction devised by Onsager for the Ising model \cite{ONSA}. 
This operator
is usually denominated the diagonal-to-diagonal transfer matrix  
and its matrix elements
are given by
\begin{equation}
\left[T_{dia}(x)\right]^{b_1,\dots,b_L}_{a_1,\dots,a_L}= \prod_{j=1}^{L} W_v(a_j,b_j|x) W_h(a_j,b_{j+1}|x) ,
\label{TDIA}
\end{equation}
where $b_{L+1}=b_1$ for periodic boundary conditions. For the  three-state scalar Potts 
model, the elements
$W_{h,v}(i,j|x)$ are the entries of the matrices defined by
Eqs.(\ref{weigh},\ref{pesos}).

The respective three-state  quantum spin chain is obtained by expanding 
the diagonal-to-diagonal transfer matrix (\ref{TDIA})
around the point $x=0$, see for instance \cite{ALC,PERK1,ALB}. At this point 
the transfer matrix reduces 
to the identity matrix and the next term in
the expansion gives rise to the Hamiltonian operator.
On a chain of $L$ sites with
periodic boundary conditions, the Hamiltonian can be written as
\begin{equation}
H^{(p)}= -\frac{2}{\sqrt{3}} \sum_{j=1}^{L-1} \left( Z_j Z^{\dagger}_{j+1}+Z^{\dagger}_{j} Z_{j+1} + X_j+X^{\dagger}_j \right ) 
-\frac{2}{\sqrt{3}} \left( Z_L Z^{\dagger}_{1}+Z^{\dagger}_{L} Z_{1} + X_L+X^{\dagger}_L\right ) , 
\label{HAMP}
\end{equation}
where the operators $Z_j$ and $X_j$ acting at the $j$-th site of the chain 
obey the $Z(3)$ algebra. Their explicit matrix representations
are,
\begin{equation}
Z=\left(
\begin{array}{ccc}
1 & 0 & 0  \\
0 & \omega & 0 \\ 
0 & 0 & \omega^2 \\ 
\end{array}
\right),~~ 
X=\left(
\begin{array}{ccc}
0 & 0 & 1  \\
1 & 0 & 0 \\ 
0 & 1 & 0 \\ 
\end{array}
\right),~~~ 
\omega=\exp(2\pi i/3) .
\end{equation}

The typical feature of an integrable quantum spin chain is that 
its spectrum 
can be parametrized 
in terms of a set of nonlinear
relations usually denominated Bethe ansatz equations 
\cite{BETHE}. This problem for
the Hamiltonian (\ref{HAMP}) was studied 
by Albertini \cite{ALB1} who has derived a Bethe equation
for the respective eigenvalues, exploring certain matrix 
identities for products
of diagonal-to-diagonal transfer matrices \cite{ALB2,BAX3}.
For comparison with our results for other
toroidal boundary conditions, we summarize in the following the main 
conclusions of the work \cite{ALB1}.  The
Hamiltonian (\ref{HAMP}) is invariant by the $Z(3)$ 
and its Hilbert space can be 
separated into three sectors labeled by
the $Z(3)$ charge  
$\exp(\frac{i\pi Q}{3})$ 
where $Q=0,1,2$.
 The
 respective Bethe equation for a given sector is \cite{ALB1}
\begin{equation}
\bigg [\frac{\sinh(\lambda_j +i\frac{\pi}{12})}{\sinh(\lambda_j -i\frac{\pi}{12})} \bigg ]^{2L}=  (-1)^{L}
\prod_{\stackrel{k=1}{k \neq j}}^{N_{Q}}
\frac{\sinh(\lambda_j-\lambda_k +i\frac{\pi}{3})}{\sinh(\lambda_j-\lambda_k -i\frac{\pi}{3})},~~j=1,\dots,N_{Q} ,
\label{BAP}
\end{equation}
where the number of Bethe roots $N_{Q}$ for each sector 
is $N_0=2L$ and $N_1=N_2=2L-2$. The  energies
$E^{(p)}_{Q}$ of the Hamiltonian (\ref{HAMP}) are expressed  in terms of the Bethe roots by,
\begin{equation}
E^{(p)}_{Q}= \sum_{j=1}^{N_{Q}} \cot\Big(\frac{\pi}{12} -i\lambda_j \Big) -\frac{2L}{\sqrt{3}} ,
\end{equation}
and several consequences of such a Bethe ansatz solution were 
studied subsequently in \cite{DAS,ALB3}. 

Before proceeding, we recall that the Hamiltonian (\ref{HAMP}) is in fact invariant 
under the larger permutation group $S_3$. This means that this operator commutes with the 
charges 
${\cal{O}}_{Z(3)}$ and ${\cal{O}}_{Z(2)}$ associated with the implementation of 
the underlying symmetries $Z(3)$ and $Z(2)$, respectively. These two charges are defined by,
\begin{equation}
{\cal{O}}_{Z(3)} = \prod_{j=1}^{L} X_j,~~ {\cal{O}}_{Z(2)}=\prod_{j=1}^{L} C_j ,
\end{equation}
where the operator $C$ corresponds to complex 
conjugation and its matrix representation is, 
\begin{equation}
C=\left(
\begin{array}{ccc}
1 & 0 & 0  \\
0 & 0 & 1 \\ 
0 & 1 & 0 \\ 
\end{array}
\right) .
\end{equation}

It turns out that the $S_3$ invariance can be broken by imposing 
suitable toroidal boundary conditions while preserving the integrability of
the corresponding quantum spin chain. This feature, for instance, can be inferred 
considering
the results for the operator content of a conformally invariant 
theory with the central charge realized by the critical three-state Potts 
quantum chain.  This model is known to have $c=\frac{4}{5}$ \cite{DOT,RIT}
and it has been argued that this class of universality is compatible
with two other toroidal boundary conditions in addition to the standard periodic
boundary \cite{CAR}.
One of the twisted boundaries  preserves the $Z(3)$ invariance, which
allows us to consider the spin at one end of the chain to differ 
by a factor $\omega^{\pm 1}$. The corresponding quantum spin chain is,
\begin{equation}
H^{(\pm)}= -\frac{2}{\sqrt{3}} \sum_{j=1}^{L-1} \left( Z_j Z^{\dagger}_{j+1}+Z^{\dagger}_{j} Z_{j+1}+ X_j+X^{\dagger}_j \right ) 
-\frac{2}{\sqrt{3}} 
\left(\frac{Z_L Z^{\dagger}_{1}}{\omega^{\pm 1}}+ \omega^{\pm 1} Z^{\dagger}_{L} Z_{1} +X_L+X^{\dagger}_L \right) . 
\label{HAMT}
\end{equation}

It is natural to expect that the spectra of the above Hamiltonians 
should also be parametrized in
terms of some transcendental set of Bethe ansatz type relations. As far as we know, this 
problem has not yet  
been considered in the literature, and in this paper, we will uncover the  respective Bethe equation.
We start by noticing that the spectra of $H^{(+)}$ and $H^{(-)}$ are related 
by complex conjugation, and therefore, we can concentrate our presentation  
to only one of the 
Hamiltonians. Here, we argue that the Bethe equation associated with 
the spectrum of the Hamiltonian $H^{(+)}$ is,
\begin{equation}
\bigg[\frac{\sinh(\lambda_j +i\frac{\pi}{12})}{\sinh(\lambda_j -i\frac{\pi}{12})} \bigg ]^{2L}=  (-1)^{L} \exp\Big[{\frac{2i \pi Q}{3}}\Big]
\prod_{\stackrel{k=1}{k \neq j}}^{\bar{N}_{Q}}
\frac{\sinh(\lambda_j-\lambda_k +i\frac{\pi}{3})}{\sinh(\lambda_j-\lambda_k -i\frac{\pi}{3})},~~j=1,\dots,\bar{N}_{Q} ,
\label{BAT}
\end{equation}
where we note the presence of an extra phase factor on the right-hand side 
of the Bethe equation 
when compared to the 
periodic case, see Eq.(\ref{BAP}). In addition to that, 
numerical study of the spectrum for small lattice sizes indicates that
the number of Bethe roots $\bar{N}_{Q}$ is now
$\bar{N}_{0}=2L-2$ and $\bar{N}_{1}=\bar{N}_{2}=2L-1$.  

If we denote by $E^{(+)}_{Q}$ the eigenvalues of
the Hamiltonian $H^{(+)}$ we find that they can be expressed in terms 
of the solutions of the Bethe
equation (\ref{BAT}) as follows,
\begin{equation}
E^{(+)}_{Q}= \sum_{j=1}^{\bar{N}_{Q}} \cot\Big(\frac{\pi}{12}-i\lambda_j\Big)+i\mu_Q -\frac{2L}{\sqrt{3}} ,
\label{ENET}
\end{equation}
where the sector dependent ``chemical potential" factor $\mu_Q$ is given by
\begin{equation}
\mu_0=0,~~\mu_1=-1,~~ \mu_2=1 .
\label{factor}
\end{equation}

The second twisted boundary corresponds to taking the complex conjugation 
of the spin at the end of the toroidal chain. This boundary breaks the 
$Z(3)$ spin rotation and the quantum spin chain only 
commutes with the charge ${\cal{O}}_{Z(2)}$. The expression of the respective
Hamiltonian is,
\begin{equation}
H^{(c)}= -\frac{2}{\sqrt{3}} \sum_{j=1}^{L-1} \left( Z_j Z^{\dagger}_{j+1}+Z^{\dagger}_{j} Z_{j+1} + X_j+X^{\dagger}_j \right ) 
-\frac{2}{\sqrt{3}} \left( Z_L Z_{1}+Z^{\dagger}_{L} Z^{\dagger}_{1} +X_L+X^{\dagger}_L \right )  
\label{HAMC}
\end{equation}

The Hilbert space of the Hamiltonian $H^{(c)}$ splits into two sectors 
associated 
with the distinct
eigenvalues $\pm 1$ of the charge ${\cal{O}}_{Z(2)}$. 
For both sectors, we find that the 
corresponding Bethe equation is,
\begin{equation}
\bigg[\frac{\sinh(\lambda_j +i\frac{\pi}{12})}{\sinh(\lambda_j -i\frac{\pi}{12})} \bigg ]^{2L}=  -(-1)^{L}
\prod_{\stackrel{k=1}{k \neq j}}^{2L}
\frac{\sinh(\lambda_j-\lambda_k +i\frac{\pi}{3})}{\sinh(\lambda_j-\lambda_k -i\frac{\pi}{3})},~~j=1,\dots,2L ,
\label{BAC}
\end{equation}
where we note the presence of 
an overall negative factor in Eq.(\ref{BAC})  compared to
the Bethe 
equation for the periodic case. 
In this case, the number of Bethe roots is fixed to $2L$ and 
the eigenvalues of the Hamiltonian $H^{(c)}$ are determined by, 
\begin{equation}
E^{(c)}= \sum_{j=1}^{2L} \cot\Big(\frac{\pi}{12}- i\lambda_j\Big) -\frac{2L}{\sqrt{3}} .
\label{ENEC}
\end{equation}

We have organized this paper as follows. In the next section, we introduce 
toroidal boundary conditions for an integrable $n$-state spin model 
by exploring a recent mapping to an equivalent solvable $n$-state  vertex 
model \cite{MAR}. This permits us to formulate twisted boundary conditions
in terms of boundary seams, responsible for changing the interactions at one end
of the respective quantum spin chains. We apply this construction
to the three-state
scalar Potts model, making it possible to build the transfer matrices 
whose Hamiltonian 
limits give rise to the quantum spin chains $H^{(\pm)}$ and $H^{(c)}$ 
discussed above. 
In sections 
3 and 4, we used certain properties of the transfer matrices 
eigenvalues to 
establish the Bethe equations parameterizing the spectra 
of the Hamiltonians 
$H^{(+)}$ and $H^{(c)}$. We investigate the solutions of the 
Bethe equations  
for $L=2$ and we use the Bethe roots to compute the Hamiltonian 
energies 
and the momenta 
of the eigenstates. The respective data for the eigenvalues of the transfer
matrix operators are also presented.

In section 5,  we use our framework to build integrable
Hamiltonians with periodic boundary conditions whose spectra are 
obtained by combining the eigenvalues of the spin chains with
different types of toroidal boundary conditions. 
In section 6,  we present our concluding remarks.
In the appendices $A$ and $B$, we summarize the analysis concerning
the completeness of the solutions of the Bethe equations for $L=3$. 

\section{Toroidal Boundary Conditions}

We start this section by describing the main results concerning 
an equivalence among 
integrable $n$-state spin
and vertex models on the square lattice \cite{MAR}. We recall that such a mapping 
assumes that
the horizontal $W_h(i,j|x)$ and vertical $W_v(i,j|x)$ 
edge weights of the spin model are
parametrized in terms of the difference of spectral 
parameters. It is well known that 
an important
ingredient for the formulation of the integrability of a given vertex model 
is the corresponding
Lax operator \cite{FAD}. This operator contains the information about the structure of the Boltzmann weights
of the vertex model denoted here by $\mathbb{L}_{i,j}^{k,l}(x) $. The indices $i,j,k,l$ represent the four state
variables assigned to the links of the square lattice meeting together at a given vertex.
In this sense, it has been argued that such 
an operator  can be 
constructed from
the spin model edge weights as follows \cite{MAR},
\begin{equation}
\label{LAX}
\mathbb{L}_{12}(x)=\sum_{i,j,k,l=1}^{n} \mathbb{L}_{i,j}^{k,l}(x)~ e_{ik} \otimes e_{jl}
=\sum_{i,j,k,l=1}^{n} W_h(j,i|x) W_v(j,k|x) \delta_{i,l}~ e_{ik} \otimes e_{jl} ,
\end{equation}
where $e_{i,j}$ denotes the standard $n \times n$ Weyl matrices and the Lax operator is  
schematically illustrated in Fig.(\ref{FigVer}).
\setlength{\unitlength}{2500sp}
\begin{figure}[H]
\begin{center}
\begin{picture}(8000,2000)
{\put(2000,900){\line(1,0){1900}}}
{\put(2960,1800){\line(0,-1){1900}}}
{\put(1900,900){\makebox(0,0){\fontsize{10}{10}\selectfont $i$}}}
{\put(4060,900){\makebox(0,0){\fontsize{10}{10}\selectfont $k$}}}
{\put(2970,-230){\makebox(0,0){\fontsize{10}{10}\selectfont $j$}}}
{\put(2990,1960){\makebox(0,0){\fontsize{10}{10}\selectfont $l$}}}
{\put(4360,870){\makebox(0,0){\fontsize{10}{10}\selectfont $=$}}}
{\put(1600,870){\makebox(0,0){\fontsize{10}{10}\selectfont $=$}}}
{\put(5870,870){\makebox(0,0){\fontsize{10}{10}\selectfont $W_v(j,k|x) W_h(j,i|x) \delta_{l,i} $}}}
{\put(930,870){\makebox(0,0){\fontsize{10}{10}\selectfont $\mathbb{L}_{i,j}^{k,l}(x)$}}}
\end{picture}
\end{center}
\caption{The illustration of the elements of the Lax operator (\ref{LAX}) on the square lattice.}
\label{FigVer}
\end{figure}

In addition to that, it has been shown that one can 
exhibit a $\mathrm{R}$-matrix
expressed also in terms of the spin edge weights, 
\begin{equation}
\label{RMA}
\mathrm{R}_{12}(x,y) =\sum_{i,j,k=1}^{n} \frac{W_h(j,i|x) W_v(j,k|x-y)}{W_h(k,i|y)} e_{ik} \otimes e_{ji} ,
\end{equation}
which together with the Lax operator satisfies the Yang-Baxter algebra \cite{BAX,FAD},
\begin{equation}
\label{RLLpure}
\mathrm{R}_{12}(x,y) \mathbb{L}_{13}(x) \mathbb{L}_{23}(y)= 
\mathbb{L}_{23}(y) \mathbb{L}_{13}(x) \mathrm{R}_{12}(x,y) .
\end{equation}

In this context, toroidal boundary conditions 
can be introduced by exploring certain internal 
symmetries of the 
Yang-Baxter algebra, see for
instance \cite{DEV}. They are associated with global 
transformations 
that leave the
Yang-Baxter algebra invariant to preserve 
integrability. This is encoded by 
a group of matrices 
$\mathrm{G}$ satisfying
the property,
\begin{equation}
[\mathrm{R}_{12}(x,y), \mathrm{G} \otimes \mathrm{G}]=0 ,
\label{prop}
\end{equation}

An immediate consequence of this symmetry is the possibility 
to define a 
family of commuting transfer matrices as follows,
\begin{equation}
T_{ver}(x)= \mathrm{Tr}_{\cal{A}} \left[\mathrm{G}_{\cal{A}} \mathbb{L}_{{\cal{A}} L}(x) \mathbb{L}_{{\cal{A}} L-1}(x) \dots \mathbb{L}_{{\cal{A}} 1}(x) \right] ,
\end{equation}
where the symbol $\cal{A}$ denotes the auxiliary space $\mathbb{C}^n$. 
We note that the periodic boundary condition corresponds to choosing $\mathrm{G}$ 
as the $n \times n$ identity matrix. The matrix elements $[T_{ver}(x)]_{a_1 \dots a_L}^{b_1 \dots b_L}$
of the transfer matrix are illustrated in Fig.(\ref{FigTVER}) where $G(a,b)$ denotes the
entries of the boundary seam $G$.
\vspace{0.5cm}
\begin{figure}[H]
\setlength{\unitlength}{1mm}
\begin{center}
\begin{picture}(55,25)
\put(-11,18){\line(1,0){47}}
\put(4,10){\line(0,1){16}}
\put(22,10){\line(0,1){16}}
%\put(13,10){\line(0,1){16}}
\put(37,17){$\cdots$}
\put(43,18){\line(1,0){48}}
\put(57,10){\line(0,1){16}}
\put(75,10){\line(0,1){16}}
%indices horizontais
{\put(-41,18){\makebox(0,0){\fontsize{12}{14}\selectfont $[T_{ver}(x)]_{a_1 \dots a_L}^{b_1 \dots b_L}$}}}
{\put(-24,18){\makebox(0,0){\fontsize{12}{14}\selectfont $=$}}}
{\put(-15,18){\makebox(0,0){\fontsize{12}{14}\selectfont $\displaystyle \sum_{c_1,\dots,c_{L+1}=1}^{n}$}}}
{\put(5,8){\makebox(0,0){\fontsize{12}{14}\selectfont $a_1$}}}
{\put(23,8){\makebox(0,0){\fontsize{12}{14}\selectfont $a_2$}}}
%{\put(14,8){\makebox(0,0){\fontsize{12}{14}\selectfont $a_3$}}}
{\put(60,8){\makebox(0,0){\fontsize{12}{14}\selectfont $a_{L-1}$}}}
{\put(76,8){\makebox(0,0){\fontsize{12}{14}\selectfont $a_{L}$}}}
%indices verticais
{\put(5,28){\makebox(0,0){\fontsize{12}{14}\selectfont $b_1$}}}
{\put(23,28){\makebox(0,0){\fontsize{12}{14}\selectfont $b_2$}}}
%{\put(14,28){\makebox(0,0){\fontsize{12}{14}\selectfont $b_3$}}}
{\put(60,28){\makebox(0,0){\fontsize{12}{14}\selectfont $b_{L-1}$}}}
{\put(76,28){\makebox(0,0){\fontsize{12}{14}\selectfont $b_{L}$}}}
%indices soma
{\put(-5,20){\makebox(0,0){\fontsize{12}{14}\selectfont $c_1$}}}
{\put(13,20){\makebox(0,0){\fontsize{12}{14}\selectfont $c_2$}}}
{\put(31,20){\makebox(0,0){\fontsize{12}{14}\selectfont $c_3$}}}
%{\put(22,20){\makebox(0,0){\fontsize{12}{14}\selectfont $c_4$}}}
{\put(49,20){\makebox(0,0){\fontsize{12}{14}\selectfont $c_{L-1}$}}}
{\put(66,20){\makebox(0,0){\fontsize{12}{14}\selectfont $c_{L}$}}}
{\put(84,20){\makebox(0,0){\fontsize{12}{14}\selectfont $c_{L+1}$}}}
{\put(101,18){\makebox(0,0){\fontsize{12}{14}\selectfont $G(c_{L+1},c_1)$}}}
\end{picture}
\end{center}
\caption{Schematic representation of the elements of the row-to-row transfer matrix of vertex models with twisted boundary $G$.}
\label{FigTVER}
\end{figure}

As usual the partition function $Z_{ver}(L)$ of the vertex model on the
$L \times L$ square lattice is given by the 
trace of the  $L$-th power of
the row-to-row transfer matrix,
\begin{equation}
\label{Zver1}
Z_{ver}(L)= \mathrm{Tr}_{{\cal V}} \left[ \Big(T_{ver}(x)\Big)^{L} \right] .
\end{equation}
where ${\cal V}= \displaystyle \prod_{j=1}^{L} \otimes {\mathrm{C}}^{n}$ is Hilbert space associated
to the  underlying quantum spin chain.

To define the corresponding quantum spin chain, we assume that 
the edge weights fulfill the 
following initial conditions,
\begin{equation}
\label{BOUND}
W_h(i,j|0)=1,~~~W_v(i,j|0)= \delta_{i,j} ,
\end{equation}
implying that at $x=0$ the Lax operator becomes the permutator $P_{12}$ on the 
tensor product $\mathbb{C}^{n} \otimes \mathbb{C}^{n}$, i.e. 
$\mathbb{L}_{12}(0)= \displaystyle \sum_{i,j=1}^{n} e_{ij} \otimes e_{ji}$. This condition also implies
that the Lax operator satisfies property (\ref{prop}) since we have $\mathbb{L}_{12}(x)=\mathrm{R}_{12}(x,0)$.

As usual, the corresponding Hamiltonian is obtained by taking the logarithmic 
derivative of the
transfer matrix at the point $x=0$ \cite{BAX,FAD}. Assuming that the boundary 
seam $\mathrm{G}$ is non-singular, we can write the Hamiltonian as,
\begin{equation}
H= \sum_{j=1}^{L-1} \mathrm{H}^{(\mathrm{bulk})}_{j,j+1} +\mathrm{H}^{(\mathrm{bound})}_{L,1} ,
\label{hamil1}
\end{equation}
where the two-body terms $\mathrm{H}^{(\mathrm{bulk})}_{j,j+1}$  and  
$\mathrm{H}^{(\mathrm{bound})}_{L1}$ are given by,
\begin{equation}
\mathrm{H}^{(\mathrm{bulk})}_{j,j+1} =P_{j,j+1} \frac{\mathrm{d}}{\mathrm{d}x} \mathbb{L}_{j,j+1}(x) \Big{|}_{x=0},~~~
\mathrm{H}^{(\mathrm{bound})}_{L1} =\mathrm{G}_{L}^{-1} \mathrm{H}^{(\mathrm{bulk})}_{L,1} \mathrm{G}_L .
\label{hamil2}
\end{equation}

At this point, we also note that $T_{ver}(0)$ plays the role of a translation operator 
which shifts the
spin generators by one site modulo $L$. More precisely, one can verify 
the following relations,  
\begin{eqnarray}
&& T_{ver}(0) \left[\mathrm{H}^{(\mathrm{bulk})}_{j,j+1} \right]\left[T_{ver}(0)\right]^{-1} =\mathrm{H}^{(\mathrm{bulk})}_{j+1,j+2},~~ j=1,\dots,L-2, \nonumber \\
&& T_{ver}(0) \left[ \mathrm{H}^{(\mathrm{bulk})}_{L-1,L} \right] \left[T_{ver}(0)\right]^{-1} =\mathrm{H}^{(\mathrm{bound})}_{L,1} .
\end{eqnarray}

Let us now discuss the above construction in the specific 
case of the three-state scalar Potts model. The first step is to solve 
Eq.(\ref{prop}) for the 
$\mathrm{R}$-matrix (\ref{RMA})
of the equivalent three-state vertex model. Starting with a generic 
$ 3 \times 3$ $\mathrm{G}$ matrix 
and considering
the structure of the edge weights (\ref{weigh}) we find that 
there exist 
three basic solutions besides the identity matrix. It turns out that any 
non-trivial solution to 
Eq.(\ref{prop}) can be seen as
a composition of the following matrices,
\begin{equation}
\mathrm{G}^{(+)}= {X}^{\dagger},~~\mathrm{G}^{(-)}=X,~~\mathrm{G}^{(c)}=C
\end{equation}
and as a result, we can define three families of commuting 
transfer matrices, 
\begin{equation}
T_{ver}^{(\pm)}(x)= \mathrm{Tr}_{{\cal{A}}} \left[\mathrm{G}^{(\pm)}_{{\cal A}} \mathbb{L}_{{\cal A}L}(x) \mathbb{L}_{{\cal A}L-1}(x) \dots \mathbb{L}_{{\cal A}1}(x) \right],~~
T_{ver}^{(c)}(x)= \mathrm{Tr}_{{\cal{A}}} \left[\mathrm{G}^{(c)}_{\cal{A}} \mathbb{L}_{{\cal A} L}(x) \mathbb{L}_{{\cal A} L-1}(x) \dots \mathbb{L}_{{\cal A} 1}(x) \right].
\end{equation}

The transfer matrices $T_{ver}^{(\pm)}(x)$ are invariant 
under the $Z(3)$ spin rotation 
and their Hamiltonian limits, apart from a constant bulk term, correspond
to the quantum spin chains $H^{(\pm)}$. Thus, we have the following commutation relations,
\begin{equation}
[T_{ver}^{(\pm)}(x),{\cal{O}}_{Z(3)}]=0,~~[T_{ver}^{(\pm)}(x),H^{(\pm)}]=0
\end{equation}
such that the spectra of $T_{ver}^{(+)}(x)$ and $T_{ver}^{(-)}(x)$ are related 
by the following equivalences,
\begin{equation}
\mathrm{Spec}\big[T_{ver}^{(+)}\big]_{Q=0}=\mathrm{Spec}\big[T_{ver}^{(-)}\big]_{Q=0},~~ 
\mathrm{Spec}\big[T_{ver}^{(+)}\big]_{Q=1}=\mathrm{Spec}\big[T_{ver}^{(-)}\big]_{Q=2},~~
\mathrm{Spec}\big[T_{ver}^{(+)}\big]_{Q=2}=\mathrm{Spec}\big[T_{ver}^{(-)}\big]_{Q=1}.
\label{spec}
\end{equation}

The transfer matrix $T_{ver}^{(c)}(x)$ keeps only the invariant 
concerning the $Z(2)$ subgroup associated with 
the charge conjugation. The respective Hamiltonian
limit gives rise to the quantum spin chain $H^{(c)}$, namely 
\begin{equation}
[T_{ver}^{(c)}(x),{\cal{O}}_{Z(2)}]=0,~~[T_{ver}^{(c)}(x),H^{(c)}]=0
\end{equation}

In the next sections, we shall see that the construction of these 
transfer matrices 
is essential  to set up the
Bethe equations for the corresponding quantum spin chains.
From the spectral relations (\ref{spec}) it is 
enough to consider the problem for $T_{ver}^{(+)}(x)$ and $T_{ver}^{(c)}(x)$.

\section{The Bethe equation for $T_{ver}^{(+)}(x)$}

Here, we follow the strategy devised by Baxter to solve
the eigenvalue problem associated with the diagonal-to-diagonal transfer matrix of the
Ising model \cite{BAX}. We recall that this kind of approach has also been  
used to uncover the Bethe equation for
the critical $Z(n)$ spin model with  periodic boundary conditions \cite{ALB1,IND}.
The first step is to build a suitable proposal for the form of the eigenvalues of
the transfer matrix $T_{ver}^{(+)}$. To this end, we observe any matrix elements of such
operator have the following structure,
\begin{equation}
\left[T_{ver}^{(+)}\right]_{a_1,\dots,a_L}^{b_1,\dots,b_L}= \Big[\frac{1}{g(x) g_1(x)}\Big]^{L} H(x)
\label{ele1}
\end{equation}
where $g(x)$ and $g_1(x)$ take into account the denominators of 
the edge weights,
\begin{equation}
g(x)=\sin(\frac{\pi}{6}+x),~~g_1(x)=\sin(\frac{\pi}{3}-x)
\end{equation}

After the denominators have been removed, each entry of the transfer
matrix is expressed in terms of the product of sines. Therefore,
the polynomial $H(x)$ can be factorized with the help
of exponentials and as a result it can be written as,
\begin{equation}
H(x)= \prod_{j=1}^{2L} \Big(c_j \exp(ix) +d_j \exp(-ix) \Big)
\label{ele2}
\end{equation}
where $c_j$ and $d_j$ are coefficients independent of $x$. 

Considering that the transfer matrix commutes 
for distinct values of the spectral parameter, we can write
the eigenvalue problem as follows,
\begin{equation}
T_{ver}^{(+)}(x) \ket{\phi} =\Lambda^{(+)}(x) \ket{\phi}
\label{TPSI}
\end{equation}
where the eigenstates $\ket{\phi}$  do not depend on $x$.

Due to the independence of the eigenstates on $x$ 
a given eigenvalue $\Lambda^{(+)}(x)$ 
can be regarded as a linear combination of the 
elements of the transfer matrix $T_{ver}^{(+)}(x)$ with constant
coefficients. As a consequence, we can conclude 
that $\Lambda^{(+)}(x)$
should have the same  form as
given by Eqs.(\ref{ele1},\ref{ele2}). To
obtain further insights on the structure of $\Lambda^{(+)}(x)$
we have studied the eigenvalue problem (\ref{TPSI}) for
small finite sizes. This analysis reveals to us that the 
the form of eigenvalues depends on the $Z(3)$ sectors
besides not given solely in terms of the product of sines.
Investigating the asymptotic behavior of $\Lambda^{(+)}(x)$ for $x \rightarrow \pm i \infty$
leads to propose 
the following form,
\begin{equation}
\Lambda^{(+)}(x)= \varrho \Big[\frac{1}{g(x) g_1(x)}\Big]^{L} \exp\Big[i\mu_{Q}(\frac{\pi}{6}-x) \Big]\prod_{k=1}^{\bar{N}_{Q}} \sin(\xi_k-\frac{\pi}{6}+x)
\label{GAMA}
\end{equation}
where $\varrho$ is a normalization, the factor $\mu_Q$ is given by Eq.(\ref{factor}) 
and the variables $\xi_k$ are related to the 
zeros of the eigenvalues, yet to be determined.   
At this stage of the analysis, the number $\bar{N}_Q$ is bounded by $2L$. We remark that
the above structure for the eigenvalue is similar to that used for periodic boundary
conditions \cite{ALB1} except by the presence of the exponential factor in Eq.(\ref{GAMA}).

It turns out that the normalization
can be fixed by observing that the transfer matrix at $x=\frac{\pi}{6}$ reduces 
to the identity matrix. As a consequence of that, the final form for the eigenvalue is, 
\begin{equation}
\Lambda^{(+)}(x)= \Big[\frac{g(\frac{\pi}{6}) g_1(\frac{\pi}{6})}{g(x) g_1(x)}\Big]^{L} \exp\Big[i\mu_{Q}(\frac{\pi}{6}-x) 
\Big]\prod_{k=1}^{\bar{N}_{Q}} 
\frac{\sin(\xi_k-\frac{\pi}{6}+x)}{\sin(\xi_k)}
\label{GAMAT}
\end{equation}
and in this normalized form, we note that
the exponential factor can be formally interpreted as the presence of 
a zero $\xi_k$ at $\pm i \infty$ for arbitrary $L$. However, the explicit computation of the eigenvalues
requires working with variables defined on a finite interval, and in practice, we have to split  
potential asymptotic contributions from the standard ratio of sines.

The next step is to determine the set of nonlinear equations 
satisfied by the variables $\xi_k$ on a finite interval.
This task can be done by exploring 
certain matrix identity
among the products 
of transfer matrices with suitable distinct spectral parameters. Here we have been inspired by the 
matrix structure 
derived for the diagonal-to-diagonal transfer matrix of the 
three-state Chiral Potts model \cite{ALB2,BAX3}. We have found  that the transfer matrix 
$T_{ver}^{(+)}(x)$ satisfies the following identity,    
\begin{eqnarray}
T_{ver}^{(+)}(x-\frac{\pi}{3}) T_{ver}^{(+)}(x-\frac{\pi}{6}) T_{ver}^{(+)}(x) &=& T_{ver}^{(+)}(0) \bigg \{
\big[f_1(x)\big]^{L} T_{ver}^{(+)}(x-\frac{\pi}{3}) 
+\big[f_2(x)\big]^{L} T_{ver}^{(+)}(x) \nonumber \\
&+&\big[f_3(x)\big]^{L} T_{ver}^{(+)}(x+\frac{\pi}{3}) 
\bigg \}
\label{funcT}
\end{eqnarray}
where the auxiliary functions $f_1(x)$, $f_2(x)$ and $f_3(x)$ are determined by,
\begin{equation}
f_1(x)=3\tan(x) \cot(x+\frac{\pi}{6}),~~f_2(x)=3 \tan(x-\frac{\pi}{6}) \cot(x),~~f_3(x)=3\tan(x-\frac{\pi}{6}) \cot(x+\frac{\pi}{6})
\label{auxf}
\end{equation}

The way we have established the functional relation (\ref{funcT},\ref{auxf}) 
is as follows. We first solve 
analytically 
some suitable entries of this matrix relation for $L=2$ and $L=3$ allowing 
the determination of
the coefficients $f_1(x)$, $f_2(x)$ and $f_3(x)$. With the help of symbolic algebra 
packages we have verified that all the entries of the functional relation are 
indeed satisfied up to $L=5$. In this task we find convenient to transform the
trigonometric functions in terms of ratio of polynomials 
in the variable $\exp(2ix)$. Based on this checking we conjecture that
Eqs.(\ref{funcT},\ref{auxf}) should be valid for arbitrary $L$ but a systematic proof 
of that has eluded us so far.

We can turn the above matrix equation into a 
functional relation for the eigenvalues by acting on both sides 
of Eq.(\ref{funcT}) on 
a common set of eigenvectors. In this procedure
we have used the fact that the transfer matrices with different spectral 
parameters commute. 
If we choose $x$ to be any
of the zeros of a given  eigenvalue, we see that the left-hand side 
of Eq.(\ref{funcT}) vanishes when it acts on the respective eigenvector.
Now by taking $x=\frac{\pi}{6}-\xi_j$,
the right hand side of Eq.(\ref{funcT}) produces the following condition for
the eigenvalues,
\begin{equation}
[f_1(\frac{\pi}{6}-\xi_j)\big]^{L} \Lambda^{(+)}(-\frac{\pi}{6}-\xi_j) +
[f_3(\frac{\pi}{6}-\xi_j)\big]^{L} \Lambda^{(+)}(\frac{\pi}{2}-\xi_j)=0
\label{funcT1}
\end{equation}

To obtain the constraints for the variables 
$\xi_k$ we substitute 
in Eq.(\ref{funcT1}) the expressions for the eigenvalue (\ref{GAMAT}) and for  
the auxiliary
functions (\ref{auxf}). After some simplifications 
we find that 
the Bethe equation 
for these variables are given by,
\begin{equation}
\bigg[\frac{\sin(\xi_j -\frac{\pi}{6})}{\sin(\xi_j)} \bigg ]^{2L}=  (-1)^{L} \exp\Big[{\frac{-2i \pi \mu_{Q}}{3}}\Big]
\prod_{\stackrel{k=1}{k \neq j}}^{\bar{N}_{Q}}
\frac{\sin(\xi_k-\xi_j +\frac{\pi}{3})}{\sin(\xi_k-\xi_j -\frac{\pi}{3})},~~j=1,\dots,\bar{N}_{Q}
\label{BAT1}
\end{equation}

We next recall that the energy of the Hamiltonian $H^{(+)}$ is computed 
by taking 
the first derivative of the logarithm of the eigenvalue at 
the value $x=0$. Considering the
normalization we have used for the edge weights we find, 
\begin{equation}
E_{Q}^{(+)}= -\frac{1}{\Lambda^{(+)}(0)} \frac{\mathrm{d}}{\mathrm{d}x} \Lambda^{(+)}(x) \Big{|}_{x=0}-4L\cot(\frac{\pi}{3})=
\sum_{j=1}^{\bar{N}_{Q}} \cot\Big(\frac{\pi}{6}-\xi_j\Big)+i\mu_Q -\frac{2L}{\sqrt{3}}
\label{ENET1}
\end{equation}

At this point, we 
note that Eqs.(\ref{BAT1},\ref{ENET1}) leads us to 
the expressions of
the Bethe 
equation (\ref{BAT}) and the energy eigenvalue
(\ref{ENET}) presented
in the introduction by choosing, 
\begin{equation}
\xi_j= i \lambda_j +\frac{\pi}{12}
\label{newpar}
\end{equation}

We  can also compute the momentum of a given eigenstate by 
identifying 
the shift operator $T_{ver}^{(+)}(0)$ with $\exp(-i\hat{P})$ where $\hat{P}$ denotes the
momenta operator. If we write the eigenvalues of the momenta as $\frac{2 \pi s_p}{L}$,
the values for the spin $s_p$ in terms of the Bethe roots are given by,
\begin{eqnarray}
s_p=\frac{i L}{2 \pi} \ln\big[\Lambda^{(+)}(0)\big] &=& \frac{i L}{2 \pi} 
\sum_{k=1}^{\bar{N}_{Q}} \ln \bigg[
\frac{\sin(\xi_k-\frac{\pi}{6})}{\sin(\xi_k)} \bigg] -\frac{L}{12}\mu_Q \nonumber \\
&=& \frac{i L}{2 \pi}    \sum_{k=1}^{\bar{N}_{Q}} \ln \bigg[
\frac{\sinh(\lambda_k+i\frac{\pi}{12})}{\sinh(\lambda_k-i\frac{\pi}{12})} \bigg] -\frac{L }{12} \mu_Q
\label{MOMET}
\end{eqnarray}

\subsection{The Bethe roots for $L=2$}

We now describe the set of roots solving the 
Bethe equation (\ref{BAT}) associated with 
the energy eigenvalues 
of the Hamiltonian $H^{(+)}$ for $L=2$. We obtain some
insight on the nature of the roots, exploring the fact 
that the eigenvectors of $T_{ver}^{(+)}(x)$ are independent of
the spectral parameter. 
By choosing a suitable value for $x$ we can compute
numerically the eigenvectors making it possible the determination
of the corresponding eigenvalues as polynomial functions on the variable $\exp(2ix)$. The zeros
of such polynomials work as starting values for the numerical 
solution of the Bethe equation (\ref{BAT}) providing  
us the Bethe
roots with high precision.

In table (\ref{tab1L2}) we have summarized our results 
for $L=2$ considering
the total number of nine eigenstates.
For each energy of the spectrum of $H^{(+)}$ 
we exhibit the sector $Q$, the spin $s_p$ of the state and the 
corresponding roots $\{\lambda_j\}$ solving the Bethe equation (\ref{BAT}). 
The energy values are obtained with the 
help of Eqs.(\ref{ENET},\ref{factor}) and we have verified that they
agree with those obtained 
from the exact diagonalization of the
Hamiltonian $H^{(+)}$. We remark that these Bethe roots allows us to compute the corresponding transfer matrix
eigenvalues with the the help of Eqs.(\ref{GAMAT},\ref{newpar}) for arbitrary values of the spectral parameter $x$. 
As an example we have added  
in the last column of table (\ref{tab1L2})
the eigenvalues 
$\Lambda^{(+)}(x)$ at the point $x=\frac{3}{5}$ reproducing
the values attained from 
the diagonalization of the operator
$T_{ver}^{(+)}(\frac{3}{5})$.
From table (\ref{tab1L2})
we note that if
a given energy in sector $Q=1$ is parametrized
by a set of roots $\{\lambda_j\}$ the same energy is obtained in sector $Q=2$ 
using the opposite set of roots $\{-\lambda_j\}$. From our solution, we see that 
this property is valid 
for arbitrary $L$ in accordance with the fact that the Hamiltonian spectra
for sectors $Q=1,2$ are the same 
since besides the spectral 
relation (\ref{spec}) we have that the invariance under reflection
maps $H^{(+)}$ 
to $H^{(-)}$.
We note that such spectral degeneracy  does not apply
to the transfer matrix eigenvalues due to the fact that  we only have the
spectral equivalence (\ref{spec}) for generic values of the spectral parameter.
In addition, 
from Eqs.(\ref{factor},\ref{MOMET}) we see that the momenta 
of the
states with roots $\{\lambda_j\}$ and $\{-\lambda_j\}$ are however 
opposite to each other.
\begin{table}[H]
\begin{center}
\begin{tabular}{|c|c|c|c|c|} \hline
$ E^{(+)} $  & $Q$ & $s_p$ & $\lambda_j$ & $\Lambda^{(+)}(\frac{3}{5})$  \\  \hline
-4.93624921    & 0 &  0 & $ \pm 0.53202156i$ & $ 0.95403776 $ \\ \hline 
-2.30940107  & 1 &  $-\frac{1}{3}$ &  $ -0.06315064 \pm 0.54254935i, ~0.19822180 $ & $\begin{array}[c]{c} 1.16822213  \\  +0.04919793i \end{array}$  \\ \hline 
-2.30940107  & 2 &  $\frac{1}{3}$ &  $ 0.06315064 \pm 0.54254935i$,~$-0.19822180 $ & $\begin{array}[c]{c} 1.16822213  \\  -0.04919793i \end{array}$  \\ \hline 
-0.84529946    & 1 &  $ \frac{2}{3} $  &  $0.16532518 \pm 0.50188873i$,~$-0.10277717 $  &  $\begin{array}[c]{c}  1.32149657 \\ -0.07557454i \end{array}$  \\ \hline 
-0.84529946    & 2 &  $-\frac{2}{3} $  &  $-0.16532518 \pm 0.50188873i,~0.10277717 $ &  $\begin{array}[c]{c}  1.32149657 \\ +0.07557454i \end{array}$   \\ \hline 
1.15470053   & 0 &  1 & $ 0,~\frac{i\pi}{2}  $  & $ 1.54404018 $ \\ \hline 
3.15470053   & 1 &  $\frac{2}{3}$ &  $ 0.24538017$,~$-0.02638725$,~$-0.64959867 $ &  $\begin{array}[c]{c} 1.79532310  \\ +0.025796612i \end{array}$ \\ \hline 
3.15470053   & 2 &  $-\frac{2}{3}$ &   $-0.24538017$,~$0.02638725,~0.64959867 $ &  $\begin{array}[c]{c} 1.79532310  \\ -0.025796612i \end{array}$  \\ \hline 
3.78154867   & 0 &  0 &  $\pm 0.12258177 $ &  $ 1.88075838 $ \\ \hline 
\end{tabular}
\end{center}
\caption{The eigenvalues ($E^{(+)}$) of the Hamiltonian $H^{(+)}$ (\ref{HAMT}) for $L=2$. 
For each energy, we indicate  
the  sector ($Q$), the spin of the state $(s_p)$ and the
values of the Bethe roots ($\lambda_j$). We have also added the respective transfer matrix eigenvalue $\Lambda^{(+)}(x)$ at $x=\frac{3}{5}$.}
\label{tab1L2}
\end{table}

Inspecting table (\ref{tab1L2})  we see 
the occurrence of low-lying eigenstates whose values for the spin
are
$\pm \frac{1}{3}$ 
and $\pm \frac{2}{3}$. 
We shall argue that 
this feature
is in accordance with the operator 
content proposed for 
the critical three-state Potts model with a $Z(3)$ invariant 
twisted boundary condition \cite{RIT,CAR}. We first recall that
the primary fields  
present in this critical system determine the 
asymptotic finite-size behaviour of the energies and the
momenta of the low-lying excitations \cite{BLO,CARD1}. For 
a given primary field
with scaling dimension 
$X=\Delta+\bar{\Delta}$ and spin 
$S=\Delta-\bar{\Delta}$ we have,   
\begin{eqnarray}
&& E(L) =e_{\infty}L +\frac{2 \pi v_F}{L} \big( \Delta+\bar{\Delta} -\frac{c}{12} +n_1+n_2\big)+{\mathcal{O}}(1/L), \nonumber \\
&& P(L)=\frac{2 \pi}{L} \big( \Delta-\bar{\Delta} +n_1-n_2\big),~~n_1,n_2=0,1,2,\dots,
\end{eqnarray}
where $e_{\infty}$ is the ground state energy in the thermodynamic limit and $v_F$ is the 
Fermi velocity of the low-lying excitations \cite{ALB1,ALB3}.
It turns out that for the $Z(3)$ twisted three-state Potts model, the corresponding 
conformal weights $(\Delta,\bar{\Delta})$ are \cite{RIT,CAR}, 
\begin{eqnarray}
&& (\Delta,\bar{\Delta})\big|_{Q=0}=\big(\frac{1}{15},\frac{1}{15}\big),~\big(\frac{2}{3},\frac{2}{3}\big) \nonumber \\
&& (\Delta,\bar{\Delta})\big|_{Q=1}=\big(\frac{1}{15},\frac{2}{5}\big),~\big(\frac{2}{3},0\big),~\big(\frac{1}{15},\frac{7}{5}\big),~\big(\frac{2}{3},3\big) \nonumber \\
&& (\Delta,\bar{\Delta})\big|_{Q=2}=\big(\frac{2}{5},\frac{1}{15}\big),~\big(0,\frac{2}{3}\big),~\big(\frac{7}{5},\frac{1}{15}\big),~\big(3,\frac{2}{3}\big)
\end{eqnarray}

From this operator content, we see that 
second lowest energy state is expected to be associated with
$X=\frac{7}{15}$  while the 
third energy state should have $X=\frac{2}{3}$, being both
doubly degenerate states since they occur in 
the sectors $Q=1,2$.
The conformal weights associated with 
the second scaling dimension
are $(\frac{1}{15},\frac{2}{5})\oplus 
(\frac{2}{5},\frac{1}{15})$ and the conformal spins of 
such states 
are therefore $S=\mp \frac{1}{3}$. For the third 
scaling dimension, 
the conformal weights 
are $(\frac{2}{3},0)\oplus (0,\frac{2}{3})$ 
and hence these states have
conformal spins $S=\pm \frac{2}{3}$. 
Considering table (\ref{tab1L2})
we conclude that  
such conformal spins indeed coincide
with the values $s_p$
obtained from the Bethe roots
for the second and the third double 
degenerate energies of the Hamiltonian spectrum.

We have performed the above 
finite size computations for  $L=3$ 
and the results are presented in Appendix B. 
This analysis confirms the   
conclusions 
we have made for the momenta of the states 
discussed above.
We have also investigated the nature
of the Bethe roots for several low-lying states with 
lattice size
$4 \leq L \leq 7$ and our results  indicate 
that the number 
of the Bethe roots for each sector should be,
\begin{equation}
\bar{N}_{0}=2L-2,~~ \bar{N}_{1}=\bar{N}_{2}=2L-1  
\end{equation}

\section{The Bethe equation for $T_{ver}^{(c)}(x)$}

Here, we use the same procedure explained in the previous section 
to determine the parameterization of the spectrum
of Hamiltonian $H^{(c)}$ in terms of a Bethe ansatz equation. We therefore avoid the 
repetition of 
technical details and concentrate  our presentation to 
the main results.
In particular, we find out that 
the structure of the eigenvalues of $T_{ver}^{(c)}(x)$ can be given as follows,
\begin{equation}
\Lambda^{(c)}(x)= \Big[\frac{g(\frac{\pi}{6}) g_1(\frac{\pi}{6})}{g(x) g_1(x)}\Big]^{L} 
\prod_{k=1}^{2L} 
\frac{\sin(\xi_k-\frac{\pi}{6}+x)}{\sin(\xi_k)}
\label{GAMAC}
\end{equation}
where now the number of sine terms is the same 
as that present in the general form (\ref{ele2}) for
the transfer matrix elements. We arrive at this conclusion by inspecting the
asymptotic limit of the eigenvalues of $T_{ver}^{(c)}(x)$ at the points $\pm i \infty$ 
for small lattice sizes.

As before, the Bethe equation for the variables $\xi_k$ is uncovered with
the help of a functional
relation for the eigenvalue involving distinct spectral parameters. This equation 
is derived from a  matrix identity for the product
of transfer matrices whose main structure is similar to that presented in 
previous section, see Eq.(\ref{funcT}). More specifically, we find that $T_{ver}^{(c)}(x)$ 
satisfies the
following matrix identity, 
\begin{eqnarray}
T_{ver}^{(c)}(x-\frac{\pi}{3}) T_{ver}^{(c)}(x-\frac{\pi}{6}) T_{ver}^{(c)}(x) &=& T_{ver}^{(c)}(0) \bigg \{
\big[f_1(x)\big]^{L} T_{ver}^{(c)}(x-\frac{\pi}{3}) 
+\big[f_2(x)\big]^{L} T_{ver}^{(c)}(x) \nonumber \\
&-&\big[f_3(x)\big]^{L} T{ver}^{(c)}(x+\frac{\pi}{3}) 
\bigg \}
\label{funcC}
\end{eqnarray}
where the main difference with Eq.(\ref{funcT}) is the  presence 
of a minus sign on the term proportional to the amplitude $f_3(x)$. We have established this relation using the
same procedure explained in the previous section. We have checked that such matrix relation is satisfied up 
to $L=5$ and we conjecture that it is valid for arbitrary $L$. 

If we choose $x=\frac{\pi}{6}-\xi_k$ the above matrix 
identity can turn into 
a constraint 
for the eigenvalues (\ref{GAMAC}) producing a set of relations for the
variables $\xi_k$. We find that such a Bethe equation is given by,
\begin{equation}
\bigg[\frac{\sin(\xi_j -\frac{\pi}{6})}{\sin(\xi_j)} \bigg ]^{2L}=  (-1)^{L+1}
\prod_{\stackrel{k=1}{k \neq j}}^{2L}
\frac{\sin(\xi_k-\xi_j +\frac{\pi}{3})}{\sin(\xi_k-\xi_j -\frac{\pi}{3})},~~j=1,\dots,2L
\label{BAC1}
\end{equation}
which becomes Eq.(\ref{BAC}) when we once again set 
$\xi_j= i \lambda_j +\frac{\pi}{12}$.

From the expression for the eigenvalue (\ref{GAMAC}), we can compute both 
the energy and the
momenta of a given state.  The energy $E^{(c)}$ is given by,
\begin{eqnarray}
E^{(c)}= -\frac{1}{\Lambda^{(c)}(0)} \frac{\mathrm{d}}{\mathrm{d}x} \Lambda^{(c)}(x) \Big{|}_{x=0}-4L\cot(\frac{\pi}{3}) & =&
\sum_{j=1}^{2L} \cot\Big(\frac{\pi}{6}-\xi_j\Big) -\frac{2L}{\sqrt{3}} \nonumber \\
&=& \sum_{j=1}^{2L} \cot\Big(\frac{\pi}{12}-i\lambda_j\Big) -\frac{2L}{\sqrt{3}} ,
\label{ENEC1}
\end{eqnarray}
while the spin $s_p$ of the eigenstate is,
\begin{equation}
s_p=\frac{i L}{2 \pi} \ln\big[\Lambda^{(c)}(0)\big] = \frac{i L}{2 \pi} 
\sum_{k=1}^{2L} \ln \bigg[
\frac{\sin(\xi_k-\frac{\pi}{6})}{\sin(\xi_k)} \bigg] 
= \frac{i L}{2 \pi}    \sum_{k=1}^{2L} \ln \bigg[
\frac{\sinh(\lambda_k+i\frac{\pi}{12})}{\sinh(\lambda_k-i\frac{\pi}{12})} \bigg] .
\label{MOMEC}
\end{equation}

\subsection{Bethe roots for L=2}

We now proceed to investigate the Bethe equation parameterization 
we have found 
for the spectrum 
of the Hamiltonian $H^{(c)}$ with $L=2$. This Hamiltonian has a $Z(2)$ 
invariance 
and the
Hilbert space splits into two sectors labeled by the 
eigenvalues of the charge 
${\cal{O}}_{Z(2)}$.  If we denote these eigenvalues by the symbol 
$\nu_c$ we find that
there are $(3^L+1)/2$ states with $\nu_c=+1$ and $(3^L-1)/2$ 
states with $\nu_c=-1$. We obtain the initial values for the 
Bethe roots 
by computing the eigenvalues of the  
transfer matrix $T_{ver}^{(c)}(x)$ as meromorphic functions on $x$. We use
this data to solve the Bethe equation (\ref{BAC}) with high 
numerical precision and the energy and the spin of the states are calculated 
with the help of Eqs.(\ref{ENEC},\ref{MOMEC}). The energies 
obtained 
from the Bethe roots are then checked with those coming from 
the exact diagonalization 
of the Hamiltonian $H^{(c)}$. The results are shown in table (\ref{tab2L2}). In this table we have also presented 
the data for the corresponding
transfer matrix eigenvalue $\Lambda^{(c)}(x)$ at $x=\frac{3}{5}$ computed  by using Eqs.(\ref{newpar},\ref{GAMAC}). We stress that 
these transfer 
matrix eigenvalues agree with the values obtained from the diagonalization of the operator $T_{ver}^{(c)}(\frac{3}{5})$.
\begin{table}[H]
\begin{center}
\begin{tabular}{|c|c|c|c|c|} \hline
$ E^{(c)} $  & $\nu_c$ & $p$ & $\lambda_j$ & $ \Lambda^{(c)}(\frac{3}{5})$ \\  \hline
-5.77350269    & +1 &  0 & $ \begin{array}[c]{c} 0.13588376 \pm 0.57005210i, \\ -0.13588376 \pm 0.57005210i \end{array}$ & $ 0.90168671$ \\ \hline 
-3.89397505 & +1 &  0 &  $ \pm 0.50828016i ,~\pm 0.524244017$ & $ 1.02818296$ \\ \hline 
-1.67372658  & $-1$ &  $\frac{1}{2}$ & $ \begin{array}[c]{c} 0.13542121 \pm 0.5300866i,\\ -0.14584486,-0.12499756+i\frac{\pi}{2} \end{array} $ & 
$\begin{array}[c]{c} 1.23423027 \\ -0.06662064i \end{array} $ \\ \hline 
-1.67372658  & $-1$ &  $-\frac{1}{2}$ &  $ \begin{array}[c]{c}  -0.13542121 \pm 0.5300866i, \\ 0.14584486,~0.12499756+i\frac{\pi}{2} \end{array} $ &  
$\begin{array}[c]{c} 1.23423027 \\ +0.06662064i \end{array} $ \\ \hline 
1.154700538  & +1 &  $1 $  &  $ \begin{array}[c]{c} -0.28844398 \pm 0.61762004i, \\ 0.55876630,~0.01812166 \end{array}$ & 
$\begin{array}[c]{c} 1.54404018 \\ +0.03665078i \end{array} $ \\ \hline 
1.154700538  & +1 &  $-1 $  &  $ \begin{array}[c]{c} 0.28844398 \pm 0.61762004i, \\ -0.55876630,~-0.01812166 \end{array} $ & 
$\begin{array}[c]{c} 1.54404018 \\ -0.03665078i \end{array} $ \\ \hline 
2.739274521  & +1 &  $0$ &  $ \pm 1.08877871i$,~$\pm 0.14111440 $ & $1.73331161 $\\ \hline 
3.983127663  & $-1$ &  $\frac{1}{2}$ &   $ \begin{array}[c]{c} -0.59441775,~-0.05108087,\\ ~0.19398856,~0.45151005+i\frac{\pi}{2} \end{array} $ &  
$\begin{array}[c]{c}  1.91132869 \\ +0.00914203i  \end{array} $ \\ \hline 
3.983127663  & $-1$ &  $-\frac{1}{2}$ &   $ \begin{array}[c]{c} 0.59441775$,~$0.05108087,\\ -0.19398856,~-0.45151005+i\frac{\pi}{2} \end{array} $ &  
$\begin{array}[c]{c}  1.91132869 \\ -0.00914203i  \end{array} $ \\ \hline 
\end{tabular}
\end{center}
\caption{The eigenvalues ($E^{(c)}$) of the Hamiltonian $H^{(c)}$ (\ref{HAMC}) for $L=2$. 
For each energy, we indicate  
the  sector ($\nu_c$), the spin of the state $(s_p)$ and the
values of the Bethe roots ($\lambda_j$).
We have also added the respective transfer matrix eigenvalue $\Lambda^{(c)}(x)$ at $x=\frac{3}{5}$.}
\label{tab2L2}
\end{table}

Interestingly enough, we observe that there are excitations 
with 
spin $\pm \frac{1}{2}$ in the charge sector with
negative eigenvalue. This is in accordance with the 
operator content 
predicted 
for the three-state Potts model with a $Z(2)$ twisted boundary condition.
In this case, the expected conformal weights are \cite{CAR},
\begin{eqnarray}
&& (\Delta,\bar{\Delta})\big|_{\nu_c=+1}=\big(\frac{1}{40},\frac{1}{40}\big),~\big(\frac{1}{8},\frac{1}{8}\big),~\big(\frac{21}{40},\frac{21}{40}\big),~\big(\frac{13}{8},\frac{13}{8}\big) \nonumber \\
&& (\Delta,\bar{\Delta})\big|_{\nu_c=-1}=\big(\frac{21}{40},\frac{1}{40}\big),~\big(\frac{1}{40},\frac{21}{40}\big),~\big(\frac{13}{8},\frac{1}{8}\big),~\big(\frac{13}{8},\frac{1}{8}\big) .
\end{eqnarray}

Considering this operator content, we note that the 
third lowest energy state is expected to have scaling dimension $X=\frac{11}{20}$.
The respective conformal weights are
$(\frac{21}{40},\frac{1}{40})\oplus(\frac{1}{40},\frac{21}{40})$ 
and this means that 
the conformal spins
are $\pm \frac{1}{2}$. From table (\ref{tab2L2}) we see that 
these are exactly the  spin values of 
the double degenerate third energy eigenstate of the Hamiltonian 
occurring in the 
sector $\nu_c=-1$. 
In appendix $C$ we have presented the Bethe roots data concerning the 
completeness of the spectrum for $L=3$. Once again, we observe that 
such low-lying excitation has half-integer spin, and this feature 
has also been checked up to $L=7$.

\section{Other families of integrable Hamiltonians}  

We can explore the invariance of the Yang-Baxter algebra (\ref{prop}) to construct
more general 
families of commuting transfer matrices. For example, we can write,
\begin{equation}
\tilde{T}_{ver}(x)= \mathrm{Tr}_{\cal{A}} \left[\mathrm{G}_{\cal{A}} \mathbb{L}_{{\cal{A}} L}(x)  \mathrm{G}_{\cal{A}} \mathbb{L}_{{\cal{A}} L-1}(x) \dots \mathrm{G}_{\cal{A}} \mathbb{L}_{{\cal{A}} 1}(x) \right] ,
\end{equation}
where the boundary seams are spread throughout the bulk by acting them
on each of the Lax operators. 

The Hamiltonian of the corresponding quantum spin chain is obtained 
as a logarithmic derivative of the 
transfer matrix at point $x=0$. As a result, we have,
\begin{equation}
\tilde{H}= \sum_{j=1}^{L} \mathrm{G}^{-1}_{j} \Bigg[
P_{j,j+1} \frac{\mathrm{d}}{\mathrm{d}x} \mathbb{L}_{j,j+1}(x) \Big{|}_{x=0} \Bigg] \mathrm{G}_{j}
\label{hamilB}
\end{equation}
where the periodic boundary condition $L+1 \equiv 1$ is assumed.

Let us apply the above construction for the classical three-state 
scalar Potts spin model. If we take the boundary matrix 
$\mathrm{G}= X^{\dagger}$ one finds that the Hamiltonian 
is given by,
\begin{equation}
\tilde{H}_1= -\frac{2}{\sqrt{3}} \sum_{j=1}^{L} \left( \frac{1}{\omega} Z_j Z^{\dagger}_{j+1}+\omega Z^{\dagger}_{j} Z_{j+1}+ X_j+X^{\dagger}_j \right ) .
\label{HAMG4}
\end{equation}

It turns out that the bulk couplings $\omega^{\pm 1}$ in the Hamiltonian (\ref{HAMG4}) 
can  be transformed
into the twisted boundary 
conditions depending on the size of the lattice. In fact, by performing the following 
local similarity transformation,
\begin{equation}
Z_j \rightarrow \omega^{-j} Z_j,~~ 
Z^{\dagger}_j \rightarrow \omega^j Z^{\dagger}_j,~~X_j \rightarrow X_j ,
\end{equation}
the Hamiltonian (\ref{HAMG4}) takes the form,
\begin{equation}
\tilde{H}_1= -\frac{2}{\sqrt{3}} \sum_{j=1}^{L-1} \left( Z_j Z^{\dagger}_{j+1}+Z^{\dagger}_{j} Z_{j+1}+ X_j+X^{\dagger}_j \right ) 
-\frac{2}{\sqrt{3}} 
\left(\frac{Z_L Z^{\dagger}_{1}}{\omega^{L}}+ \omega^{L} Z^{\dagger}_{L} Z_{1} +X_L+X^{\dagger}_L \right) . 
\label{HAMG4T}
\end{equation}

We note that  the boundary term of the Hamiltonian (\ref{HAMG4T}) 
depends on the lattice size modulo 3. 
For $L=3m$, where $m$ is a positive integer, the twisted term becomes unity, and we have periodic 
boundary conditions, but when
$L=3m \pm 1$ we have twisted boundary conditions associated 
with the matrices $\mathrm{G}^{(\pm)}$. Therefore, the spectrum of the
Hamiltonian (\ref{HAMG4}) can be obtained from the spectral relations,
\begin{equation}
\mathrm{Spec}\Big[\tilde{H}_1\Big]_{L=3m}= \mathrm{Spec}\Big[H^{(p)}\Big]_{L=3m},~~
\mathrm{Spec}\Big[\tilde{H}_1\Big]_{L=3m \pm 1}= \mathrm{Spec}\Big[H^{(\pm)}\Big]_{L=3m \pm 1} .
\end{equation}

We now consider the situation in which the boundary matrix  is $\mathrm{G}=C$. In this
case the corresponding Hamiltonian is,
\begin{equation}
\tilde{H}_2= -\frac{2}{\sqrt{3}} \sum_{j=1}^{L} \left( Z_j Z_{j+1}+Z^{\dagger}_{j} Z^{\dagger}_{j+1}+ X_j+X^{\dagger}_j \right ) .
\label{HAMCC}
\end{equation}

Once again, the eigenvalues of the Hamiltonian (\ref{HAMCC}) can be determined in terms of the spectrum 
of the three-state Potts quantum chain with boundary conditions depending on the lattice size. This can be seen by considering 
the charge conjugation 
symmetry only for even sites, namely
\begin{equation}
Z_j \rightarrow Z^{\dagger}_j,~~ 
Z^{\dagger}_j \rightarrow Z_j,~~X_j \rightarrow X^{\dagger}_j,~~
X^{\dagger}_j \rightarrow  X_j~~\mathrm{for}~~j=2,4,6, \dots.
\label{TRAZ2}
\end{equation}

By applying this transformation, we find that the Hamiltonian (\ref{HAMCC}) changes to,
\begin{equation}
\tilde{H}_2= -\frac{2}{\sqrt{3}} \sum_{j=1}^{L-1} \left( Z_j Z^{\dagger}_{j+1}+Z^{\dagger}_{j} Z_{j+1}+ X_j+X^{\dagger}_j \right ) +\tilde{H}^{(\mathrm{bound})}_2
\label{HAMCCC}
\end{equation}
where the boundary term $\tilde{H}^{(bound)}_2$ depends on the parity of the lattice size,
\begin{equation}
\tilde{H}^{(\mathrm{bound})}_2= \Bigg \{ 
\begin{array}{c} -\frac{2}{\sqrt{3}} \left( Z_L Z^{\dagger}_{1}+Z^{\dagger}_{L} Z_{1}+ X_L+X^{\dagger}_L \right ) ~~\mathrm{for}~~L \in \mathrm{even} \\
-\frac{2}{\sqrt{3}} \left( Z_L Z_{1}+Z^{\dagger}_{L} Z^{\dagger}_{1}+ X_L+X^{\dagger}_L \right ) ~~\mathrm{for}~~L \in \mathrm{odd}, 
\end{array}
\label{HAMCCC1}
\end{equation}
and as a result, we have the following spectral equivalence,
\begin{equation}
\mathrm{Spec}\Big[\tilde{H}_2\Big]_{L=\mathrm{even}}= \mathrm{Spec}\Big[H^{(p)}\Big]_{L=\mathrm{even}},~~
\mathrm{Spec}\Big[\tilde{H}_2\Big]_{L=\mathrm{odd}}= \mathrm{Spec}\Big[H^{(c)}\Big]_{L=\mathrm{odd}}.
\end{equation}

The  quantum spin chain (\ref{HAMCC}) appears to be an interesting integrable 
variant of the 
three-state Potts spin chain
with periodic boundary conditions. A nice feature of such spin chain is that 
it provides a lattice
realization of the complete set of the lowest weights of the Virasoro minimal model with 
central charge $c=4/5$. To this end let us denote by $\phi_{r,s}$ the primary fields of
this theory whose conformal weights are given by \cite{DOT},
\begin{equation}
\Delta_{r,s}= \frac{(6r-5s)^2-1}{120},~~1\leq r \leq 2,~~1 \leq s \leq 5
\label{KAC}
\end{equation}
where the underlying operator algebra is invariant under a $Z(2)$ invariance of the
form \cite{ZAMOZAMO},
\begin{equation}
\phi_{r,s} \rightarrow (-1)^{s+1} \phi_{r,s}
\label{SIMZ2}
\end{equation}

For $L$ even the spectrum of $\tilde{H}_2$ 
is the same of the three-state Potts spin chain with periodic boundary conditions 
whose operator content
was discussed in \cite{DOT,RIT}. The
underlying conformal weights correspond to  
that of the primary 
fields which are even under the $Z(2)$ invariance (\ref{SIMZ2}),
\begin{equation}
\Delta_{1,3}=\frac{2}{3},~\Delta_{1,5}=3,~\Delta_{2,1}=\frac{2}{5},~\Delta_{2,3}=\frac{1}{15},~\Delta_{2,5}=\frac{7}{5}
\end{equation}
but when $L$ is odd the spectrum of $\tilde{H}_2$ coincides with that of the three-state Potts spin chain 
with a $Z(2)$ twisted boundary. The operator content was proposed in \cite{CAR} and the respective 
conformal weights correspond to those of the primary fields 
which are odd under the $Z(2)$ invariance (\ref{SIMZ2}),
\begin{equation}
\Delta_{1,2}=\frac{1}{8},~\Delta_{1,4}=\frac{13}{8},~\Delta_{2,2}=\frac{1}{40},~\Delta_{2,4}=\frac{21}{40}
\end{equation}
and altogether give the values of the lowest weights predicted by Eq.(\ref{KAC}).

\section{Conclusions}

The main goal of this paper was to establish the Bethe type 
equations
for the spectra of the critical 
three-state Potts quantum spin chain with twisted boundary 
conditions. To accomplish this task,  we  first constructed 
the respective transfer matrices exploring a mapping among
integrable classical spin and vertex models. This equivalence
permits us to formulate a toroidal boundary condition in terms
of a boundary seam responsible for changing the interaction among 
the spins
at one end of the chain. Exploring the internal symmetries 
of the Yang-Baxter
algebra we find that there are two other classes of integrable 
twisted boundaries
besides the periodic boundary condition.

The first class of twisted boundary preserves 
the  $Z(3)$ invariance 
of the three-state Potts model. In this case, the ansatz 
for the transfer matrix 
eigenvalue
needs a suitable  modification compared with that devised for the
periodic boundary condition \cite{ALB1}. The second class 
of twisted boundary 
keeps only
the $Z(2)$ invariance and the transfer 
matrix eigenvalue has the same form of the periodic case,
but now the number of the
Bethe roots is fixed. In both cases, the corresponding Bethe
equations are found by exploring certain matrix identities for
the product of transfer matrices. The structure of
these relations are similar for both twisted
boundaries except for a sign in one of the amplitudes 
of the linear combination of transfer matrices.
The basic form of the Bethe equations is that found for
the periodic boundary condition except by the presence 
of phase factors directly
related to the ratio of the eigenvalues of the matrices
representing the boundary seams. In this sense, the 
situation is equivalent
to that found for integrable vertex systems invariant 
under continuous
symmetries, such as the six-vertex model \cite{BAX,FMM}. 
We have studied all the eigenvalues of the twisted three-state
Potts spin chain for $L=2,3$ in terms of the root content of
the corresponding Bethe equations. This allowed us to compute
the momenta of the eigenstates and we found that some of the
low-lying excitations have fractional values for 
the respective spin. We have argued
that they agree with   
the conformal spins predicted by the corresponding conformal field
theories. 

It is conceivable that the framework discussed in this paper may be applied
to other integrable 
classical spin model, such as those invariant under the $Z(n)$ group of
discrete rotations in the spin space. Recall here that the  $Z(n)$ algebra
is defined by,
\begin{equation}
Z^{n}=X^{n}=1,~~ZX=\omega XZ,~~\omega=\exp(2\pi i/n)
\end{equation}
where the $Z$ and $X$ are $n\times n$ operators generalizing the Pauli matrices,
\begin{equation}
Z= \left( \begin{array}{cccccc}
1 & 0 & 0 & \dots & 0 & 0\\
0 & \omega & 0 & \dots & 0 & 0\\
0 & 0 & \omega^2 &  & 0 & 0\\
\vdots & \vdots & \vdots & \ddots & \vdots & \vdots\\
0 & 0 & 0 & \dots & \omega^{n-2} & 0\\
0 & 0 & 0 & \dots & 0 & \omega^{n-1}\\
\end{array} \right)_{n \times n}~~~
X= \left( \begin{array}{cccccc}
0 & 0 & 0 & \dots & 0 & 1\\
1 & 0 & 0 & \dots & 0 & 0\\
0 & 1 & 0 &  & 0 & 0\\
\vdots & \vdots & \vdots & \ddots & \vdots & \vdots\\
0 & 0 & 0 & \dots & 0 & 0\\
0 & 0 & 0 & \dots & 1 & 0\\
\end{array} \right)_{n \times n}~~~
\end{equation}

The typical example is
the Fateev-Zamolodchikov model whose edge weights are given by \cite{FAZA},
\begin{equation}
W_h(a,b|x)= \prod_{j=1}^{|a-b|} \frac{\sin \left( (2j-1) \frac{\pi}{2n} -x \right)}{\sin \left( (2j-1) \frac{\pi}{2n} +x \right)},~~
W_v(a,b|x)= \prod_{j=1}^{|a-b|} \frac{\sin \left( (j-1) \frac{\pi}{n} +x \right)}{\sin \left( \frac{j\pi}{n} -x \right)} ,
\end{equation}
and the associated quantum spin chain with periodic boundary conditions is given by \cite{PERK1,ALB},
\begin{equation}
H_n^{(p)}= -\sum_{j=1}^{L} \sum_{k=1}^{n} \frac{1}{\sin(k\pi/n)} \Big( \left[Z_{j} Z^{\dagger}_{j+1}\right]^{k} +\left[X_j\right]^{k} \Big),~~Z^{\dagger}_{L+1}= Z^{\dagger}_1
\label{FATZAMOchain}
\end{equation}

We have investigated 
the symmetries of the Yang-Baxter algebra associated with 
the equivalent vertex 
model built from the edge weights of the
Fateev-Zamolodchikov model. We find that 
there are $(n-1)$ integrable toroidal
boundary conditions compatible
with the $Z(n)$ invariance besides 
the above mentioned 
periodic boundary condition. 
The respective boundary 
seams are given by,
\begin{equation}
\mathrm{G}^{(l)}= X^{n-l},~~l=1,\dots,n-1
\end{equation}
which now leads us to the following twisted boundary conditions, 
\begin{equation}
Z^{\dagger}_{L+1}=\omega^{-l} Z^{\dagger}_1 .
\label{TOROZN}
\end{equation}

For $n \geq 3$, we also have an integrable 
twisted  boundary condition associated with 
the breaking of the 
$Z(n)$ rotational invariance of the 
spin model. In this case
the boundary seam is,
\begin{equation}
\mathrm{G}^{(c)}= \left( \begin{array}{cccccc}
1 & 0 & 0 & \dots & 0 & 0\\
0 & 0 & 0 & \dots & 0 & 1\\
0 & 0 & 0 & \dots & 1 & 0\\
\vdots & \vdots & \vdots & \scalebox{-1}[1]{$\ddots$} & \vdots & \vdots\\
0 & 0 & 1 & \dots & 0 & 0\\
0 & 1 & 0 & \dots & 0 & 0\\
\end{array} \right)_{n \times n}
\end{equation}
and the respective toroidal boundary condition 
corresponds 
to charge conjugation at the end of the chain,
\begin{equation}
Z^{\dagger}_{L+1}= Z_1 .
\label{TOROZ2}
\end{equation}

In principle, it should be possible to derive 
the Bethe equations 
for the $Z(n)$ Fateev-Zamolodichikov model with the above 
toroidal  boundary conditions. For the boundary seams 
preserving the $Z(n)$ invariance, we expect that 
the expression for the eigenvalue of the transfer matrix 
should contain
additional suitable exponential terms that multiply the 
standard product of sine
ratios. At  present, however, 
the precise determination of the dependence of the respective
phase factors
on the many possible $Z(n)$ 
sectors is an open question. Another relevant technical 
issue we have to deal with 
is the
derivation of the functional relations constraining the 
eigenvalues for
arbitrary spectral parameters.
In any case, one expects that
the structure of the Bethe equations will be dependent 
on whether $n$ is odd
or even since this is the situation for periodic 
boundary condition \cite{IND}. We hope to return to 
some of these problems in future work.

\section*{Acknowledgments}

This work was supported in part by the Brazilian Research Council CNPq 305617/2021-4.

\addcontentsline{toc}{section}{Appendix A}
\section*{\bf Appendix A: The spectrum of $H^{(+)}$ for $L=3$ }
%\label{APENA}
\setcounter{equation}{0}
\renewcommand{\theequation}{A.\arabic{equation}}

In table (\ref{tab1L3}) we present the Bethe roots
associated with 
the spectrum of the Hamiltonian $H^{(+)}$ for $L=3$ and for the sectors $Q=0,1$.
The energy eigenvalues 
for sector $Q=2$ are the same as those of sector $Q=1$. Recall that the Bethe roots
for $Q=2$ are obtained from the data of the sector $Q=1$ by reflecting the roots 
around the origin.  
\begin{table}[H]
\begin{center}
\begin{tabular}{|c|c|c|c|} \hline
$ E^{(+)} $  & $Q$ & $s_p$ & $\lambda_j$  \\  \hline
-7.99554373    & 0 &  0 & $ 0.09225911 \pm 0.53094089i$,~$-0.09225911 \pm 0.53094089i $\\ \hline 
-6.10495278  & 1 &  $-\frac{1}{3}$ &  $-0.12712369 \pm 0.54308217i,~0.04043704 \pm 0.53625306i,~0.26682758 $ \\ \hline
-4.92871242  & 1 &  $\frac{2}{3}$ &  $ 0.22896922 \pm 0.49856802i,~0.011447035 \pm 0.51625714i$,~$-0.17891715 $ \\ \hline
-2.74536016    & 0 &  $ 1 $  &  $ 0.09061055 \pm 0.52433255i $,~$-0.09183193$,~$-0.15089655 +i\frac{\pi}{2}$     \\ \hline 
-2.74536016    & 0 &  $-1 $  &  $ -0.09061055 \pm 0.52433255i $,~$0.09183193$,~$0.15089655 +i\frac{\pi}{2}$     \\ \hline 
-2.57115043    & 1 &  $ \frac{5}{3} $  &  $-0.15570094 \pm 0.53580121i$,~$0.23995930 \pm 0.50494266i$,~$0.02135245$\\ \hline 
-1.40303732   & 0 &  0 & $ \pm 0.52006713i $,~$\pm 0.18946913 $ \\ \hline 
-1.36808057   & 1 &  $\frac{2}{3}$ &  $  0.06125792 \pm  0.52659924i $,~$0.29627169$,~-$0.10619000$,~-$0.65151671+i\frac{\pi}{2}$ \\ \hline 
-1.03735236   & 1 &  $\frac{5}{3}$ &  $-0.0995242 \pm 0.52182842i $,~ $0.07059464$,~$0.31859672$,~$-0.454114613+i\frac{\pi}{2}$\\ \hline 
0.96180183  & 0 &  1 &  $-0.16995935 \pm 0.53302564i $,~$0.00861770$,~$0.19979605$ \\ \hline 
0.96180183  & 0 &  $-1$ &  $0.16995935 \pm 0.53302564i $,~$-0.00861770$,~$-0.19979605$ \\ \hline 
2.16572177  & 1 &  $-\frac{1}{3}$ &  $ 0.26012831 \pm 0.50916006i $,~$0.04253895$,~$-0.12043166 $,~$-0.86708297+i\frac{\pi}{2}$ \\ \hline 
2.47037783  & 0 &  0 &  $\pm 0.09118002$,~$\pm 1.07984698i$ \\ \hline 
3.60850280  & 1 &  $\frac{5}{3}$ &  $0.15793948 \pm 0.75208457i$,~$0.23369692$,~$0.00328566$,~$-0.19352258$ \\ \hline 
3.93923101  & 1 &  $-\frac{1}{3}$ &  $-0.21465852 \pm 0.95822988i $,~$0.32669908$,~$ 0.07393824$,~$-0.09692341$  \\ \hline 
5.24765993  & 0 &  1 &  $-0.21736842$,~-$0.028038390$,~$0.13212693$,~$0.99943937+i\frac{\pi}{2}$ \\ \hline 
5.24765993  & 0 &  $-1$ &  $0.21736842$,~$0.028038390$,~$-0.13212693$,~$-0.99943937+i\frac{\pi}{2}$ \\ \hline 
6.29679300  &  1 & $\frac{2}{3} $ & $0.38083053$,~$0.11194307$,~$-0.039031367$,~$-0.22789695$,~$0.19855140+i\frac{\pi}{2}$\\ \hline 
\end{tabular}
\end{center}
\caption{The eigenvalues ($E$) of the Hamiltonian $H^{(+)}$ (\ref{HAMT}) for $L=3$ 
together with 
respective sector ($Q$) and
the values of the Bethe roots ($\lambda_j$). The Bethe roots and the spin ($s_p$) for 
the sector $Q=2$ have the opposite 
values of those of sector $Q=1$.}
\label{tab1L3}
\end{table}

\addcontentsline{toc}{section}{Appendix B}
\section*{\bf Appendix B: The spectrum of $H^{(c)}$ for $L=3$ }
%\label{APENB}
\setcounter{equation}{0}
\renewcommand{\theequation}{B.\arabic{equation}}

In tables (\ref{tab2L3A},\ref{tab2L3B}) we present the Bethe roots
associated with 
the spectrum of the Hamiltonian $H^{(c)}$ for $L=3$. 
\begin{table}[H]
\begin{center}
\begin{tabular}{|c|c|c|c|} \hline
$ E^{(c)} $  & $\nu_c$ & $s_p$ & $\lambda_j$  \\  \hline
-8.53674848    & $+$ &  0 & $0.20734291 \pm 0.56486237i$ ,~$-0.20734291 \pm 0.56486237i$ ,~$\pm 0.54454434i$\\ \hline 
-7.28852460 & $+$ &  0 &  $ 0.08974272 \pm 0.51111514i $,~$-0.08974272 \pm 0.51111514i$,$\pm 0.58987350$ \\ \hline 
-5.60503415    & $-$ &   $\frac{1}{2}$ & $ \begin{array}[c]{c}  0.20589826 \pm 0.53508028i ,~-0.00365705 \pm 0.52532788i  
\\ ~-0.21805593,~-0.18642647+i\frac{\pi}{2}  \end{array} $ \\ \hline 
-5.60503415    & $-$ & $-\frac{1}{2}$ & $ \begin{array}[c]{c}  -0.20589826 \pm 0.53508028i ,~0.00365705 \pm 0.52532788i  
\\ ~0.21805593,~0.18642647+i\frac{\pi}{2}  \end{array} $ \\ \hline 
-2.99101531 & $+$ &  $1$ &  $\begin{array}[c]{c} 0.30208439 \pm 0.62106348i ,~0.06639032 \pm 0.53501552i \\ -0.10649764,~-0.63045180 \end{array}$ \\ \hline 
-2.99101531 & $+$ &  $-1$ &  $\begin{array}[c]{c} -0.30208439 \pm 0.62106348i ,~-0.06639032 \pm 0.53501552i \\ 0.10649764,~0.63045180 \end{array}$ \\ \hline 
-2.62684813 & $-$ &  $\frac{3}{2}$ &  $ 0.2104731 \pm 0.5274506i $,~$-0.2104731 \pm 0.5274506i$,~$0 $,~$i \frac{\pi}{2}$ \\ \hline 
-2.44884573 & $+$ &  $0$ &  $ \pm 0.52250539i$,~$\pm 1.14014422i$,~$\pm 0.21314320$\\ \hline 
-2.43937374 & $+$ &  $1$ &  $ \begin{array}[c]{c} 0.09827709 \pm 0.51654092i ,~-0.36360829 \pm 0.61547441i \\ 0.60739795,~-0.07673556 \end{array} $\\ \hline 
-2.43937374 & $+$ &  $-1$ &  $ \begin{array}[c]{c} -0.09827709 \pm 0.51654092i ,~0.36360829 \pm 0.61547441i \\ -0.60739795,~0.07673556 \end{array} $\\ \hline 
-0.91370050    & $-$ &   $\frac{1}{2}$ & $ \begin{array}[c]{c}   0.03789187 \pm 0.52617360i,~0.24829909,~-0.12516643 
\\ ~-0.65421390,~0.45529749+i\frac{\pi}{2}  \end{array} $ \\ \hline 
-0.91370050    & $-$ &   $-\frac{1}{2}$ & $ \begin{array}[c]{c}   -0.03789187 \pm 0.52617360i,~-0.24829909,~0.12516643 
\\ ~0.65421390,~-0.45529749+i\frac{\pi}{2}  \end{array} $ \\ \hline 
\end{tabular}
\end{center}
\caption{The eigenvalues ($E^{(c)}$) of the Hamiltonian $H^{(c)}$ (\ref{HAMC}) for $L=3$. 
For each energy, we indicate  
the  sector ($\nu_c$), the spin of the state $(s_p)$ and the
values of the Bethe roots ($\lambda_j$).}
\label{tab2L3A}
\end{table}

\begin{table}[H]
\begin{center}
\begin{tabular}{|c|c|c|c|} \hline
$ E^{(c)} $  & $\nu_c$ & $s_p$ & $\lambda_j$  \\  \hline
0    & $-$ &   $\frac{3}{2}$ & $ \begin{array}[c]{c}   -0.11216447 \pm 0.51884900i,~0.26708111,~0.05141273 
\\ ~-0.62612820,~0.53196329+i\frac{\pi}{2}  \end{array} $ \\ \hline 
0    & $-$ &   $-\frac{3}{2}$ & $ \begin{array}[c]{c}   0.11216447 \pm 0.51884900i,~-0.26708111,~-0.05141273 
\\ ~0.62612820,~-0.53196329+i\frac{\pi}{2}  \end{array} $ \\ \hline 
0.43937374  & $+$ &  $1 $  &  $ \begin{array}[c]{c} 0.09986159 \pm 1.07140349i,~-0.20825039 \pm 0.52974992i \\ 0.21482995,~0.00194763 \end{array} $ \\ \hline
0.43937374  & $+$ &  $-1 $  &  $ \begin{array}[c]{c} -0.09986159 \pm 1.07140349i,~0.20825039 \pm 0.52974992i \\ -0.21482995,~-0.00194763 \end{array} $ \\ \hline
2.14093253    & $-$ &   $\frac{1}{2}$ & $ \begin{array}[c]{c}   -0.2309567 \pm 0.52156409i,~0.67763416,~0.13923161 
\\ ~-0.02538219,~-0.32957012+i\frac{\pi}{2}  \end{array} $ \\ \hline 
2.14093253    & $-$ &   $-\frac{1}{2}$ & $ \begin{array}[c]{c}   0.2309567 \pm 0.52156409i,~-0.67763416,~-0.13923161 
\\ ~0.02538219,~0.32957012+i\frac{\pi}{2}  \end{array} $ \\ \hline 
2.27326872  & $+$ &  $0 $  &  $0.39260729 \pm 0.61234873i$,~$-0.39260729 \pm 0.61234873i$,~$\pm 0.08548240 $ \\ \hline
2.85105230  & $+$ &  $0$ &  $ \pm 0.75371919i$,~$\pm 0.62598793 $,~$\pm 0.09437378$\\ \hline 
3.46410161  & $-$ &  $\frac{3}{2} $  &  $\pm 0.90062802i$,~$\pm 0.21563865$,~$0$,~$i\frac{\pi}{2}$ \\ \hline 
4.37780211 & $-$ &   $\frac{1}{2}$ & $ \begin{array}[c]{c}   0.41477805 \pm 0.61080054i,~0.09748499,~-0.06279076 
\\ ~-0.28298807,~-0.58126226+i\frac{\pi}{2}  \end{array} $ \\ \hline 
4.37780211 & $-$ &   $-\frac{1}{2}$ & $ \begin{array}[c]{c}   -0.41477805 \pm 0.61080054i,~-0.09748499,~0.06279076 
\\ ~0.28298807,~0.58126226+i\frac{\pi}{2}  \end{array} $ \\ \hline 
4.99101531 & $+$ &  $1$ &  $ \begin{array}[c]{c}  -0.28591401 \pm 1.02997584i,~0.68123228,~0.14138378 \\-0.02346981,~-0.22731814 \end{array} $ \\ \hline 
4.99101531 & $+$ &  $-1$ &  $ \begin{array}[c]{c}  0.28591401 \pm 1.02997584i,~-0.68123228,~-0.14138378 \\0.02346981,~0.22731814 \end{array} $ \\ \hline 
6.09094975  & $-$ & $\frac{3}{2}$ &   $\pm 0.71310985$,~$\pm 0.16240235$,~$0$,~$i\frac{\pi}{2}$ \\ \hline 
6.22159457  & $+$ & $0$ &   $\pm 0.29687041$,~$\pm 0.07515995$,~$\pm 0.61636095+ i\frac{\pi}{2}$ \\ \hline 
\end{tabular}
\end{center}
\caption{The eigenvalues ($E^{(c)}$) of the Hamiltonian $H^{(c)}$ (\ref{HAMC}) for $L=3$. 
For each energy, we indicate  
the  sector ($\nu_c$), the spin of the state $(s_p)$ and the
values of the Bethe roots ($\lambda_j$).}
\label{tab2L3B}
\end{table}

{}

\end{document}